\documentclass[%
  twocolumn,
  oneside,
  amsmath,amssymb,
  aps,
  pra,
  nofootinbib,
  superscriptaddress,
  a4paper
]{revtex4-2}

\usepackage{graphicx}% 
\usepackage[usenames,dvipsnames]{xcolor}
\usepackage{siunitx}
\usepackage{physics}  
\usepackage{natbib}

\usepackage[
  bookmarks=false,
  colorlinks,
  linkcolor=blue,
  urlcolor=blue,
  citecolor=blue,
  plainpages=false,
  pdfpagelabels,
  final,
  breaklinks=true
]{hyperref}
\hypersetup{
pdftitle={Title}, 
pdfauthor={Emilio Pisanty}
}
\DeclareSIUnit{\au}{{a.u.}}

\def\beq#1{\begin{equation}\label{#1}}
\def\eeq{\end{equation}}

\begin{document}

\author{V.~V.~Strelkov}
\email{strelkov.v@gmail.com}
\affiliation{Prokhorov General Physics Institute of the Russian Academy of Sciences, Vavilova street 38, Moscow 119991, Russia}
\affiliation{Moscow Institute of Physics and Technology (National Research University), Dolgoprudny, Moscow Region 141700, Russia}

\author{M.~A.~Khokhlova}
\affiliation{King’s College London, Strand Campus, London WC2R 2LS, UK}
%\email{margarita.khokhlova@kcl.ac.uk}

\title{Phase-matching gating for isolated attosecond pulse generation}

%%%%%%%%%%%%%%%%%%% abstract %%%%%%%%%%%%%%%%
\begin{abstract}
 We investigate the production of an isolated attosecond pulse~(IAP) via the phase-matching gating of high-harmonic generation by intense laser pulses. Our study is based on the integration of the propagation equation for the fundamental and generated fields with nonlinear polarisation found via the numerical solution of the time-dependent Schr\"odinger equation. We study the XUV energy as a function of the propagation distance (or the medium density) and find that the onset of the IAP production corresponds to the change from linear to quadratic dependence of this energy on the propagation distance (or density). Finally, we show that the upper limit of the fundamental pulse duration for which the IAP generation is feasible is defined by the temporal spreading of the fundamental pulse during the propagation. This nonlinear spreading is defined by the difference of the group velocities for the neutral and photoionised medium. 
\end{abstract}
%%%%%%%%%%%%%%%%%%% abstract %%%%%%%%%%%%%%%%

\maketitle

\section{Introduction}
One of the central directions of attosecond physics~\cite{Corkum2007, Krausz2009, Villeneuve2018, Ryabikin2023} remains the advance of the attosecond pulses generated via the process of high-harmonic generation~(HHG), typically in the shape of an attosecond pulse train. Attosecond XUV pulses have already proven to be a unique tool, which is used to study ultrafast electronic processes in atoms, molecules and solids at their natural time scale. Among improvements of the attosecond pulses in duration, making them shorter to increase the temporal resolution of the processes, and in intensity, making them brighter to cover a wider range of processes, there is an important goal: the creation of isolated attosecond pulses~(IAPs)~\cite{Hentschel2001}. IAPs combined with intense laser pulses provide a very powerful pump-probe technique, which can be exploited for attosecond streaking~\cite{Goulielmakis2004, Schultze2010} and transient-absorption spectroscopy~\cite{Argenti2015, Gruson2016, Geneaux2019}. Recently, the high demand for the generation of IAPs was emphasised by the realisation of attosecond-pump attosecond-probe experiments~\cite{Kretschmar2023}, which rely on high XUV energies to work.

The generation of IAPs via HHG has been demonstrated using a number of gating schemes~\cite{Chini2014}: amplitude gating (where typically a few-cycle laser pulse provides the gating)~\cite{Goulielmakis2008, Xue2022, Witting2022}, polarisation gating (first demonstrated in~\cite{Zair2004} and later in~\cite{Sola2006, Sansone2006}), double optical gating~\cite{Mashiko2008, Takahashi2013}, spatio-temporal gating (the attosecond lighthouse technique~\cite{Vincenti2012, Kim2013, Hammond2016} and noncollinear gating~\cite{Kennedy2022}), and phase-matching gating~\cite{Sandhu2006, Jullien2008, phase-matching_gating, Hernandez-Garcia2017, Schotz2020} (also called ionisation gating in case of the microscopic effect). In this paper we focus on phase-matching gating.

The phase-matching gating scheme is based on the creation of a temporal window for the macroscopically-propagating HHG emission due to the compensation of the phase mismatch of contributions coming from the free electrons and from the neutral atoms~\cite{KhokhlovaStrelkov2020, Khokhlova2023}. As far as the relative contribution changes with the ionisation degree during the laser pulse, then this finite temporal window can be achieved, moreover this window can be short enough to fit a single attosecond pulse, or an IAP, only. 

In this paper we perform a numerical study of macroscopic HHG in argon gas by solving the 3D time-dependent Schr\"odinger equation~(TDSE) coupled with the reduced propagation equation~\cite{KhokhlovaStrelkov2020, Khokhlova2023}. We choose the intense laser pulse in a way that it results in the generation of an IAP within phase-matching gating. We calculate the propagation of IAP generated in gas for different laser pulse parameters~--- the pulse intensity and the pulse duration~--- and we find the optimal regime for the IAP generation.

\section{Methods}
We numerically integrate the 1D propagation equation, and at each step of the propagation we calculate the nonlinear polarisation of the medium by solving 3D TDSE for a model argon atom (for more details see~\cite{KhokhlovaStrelkov2020, Khokhlova2023}). For HHG by a spatial flat-top beam~\cite{phase-matching_gating, Constant_2012, Veyrinas2023} the laser intensity is almost constant up to a certain distance from the beam axis, so the 1D propagation equation is adequate. 

All simulations are done for an argon target with density $\SI{3e18}{cm^{-3}}$ and a Ti:Sapphire laser pulse with wavelength \SI{800}{nm} and a $\sin^2$ envelope.

\section{Results}
\begin{figure}
\centering
\includegraphics[width=1.0\linewidth]{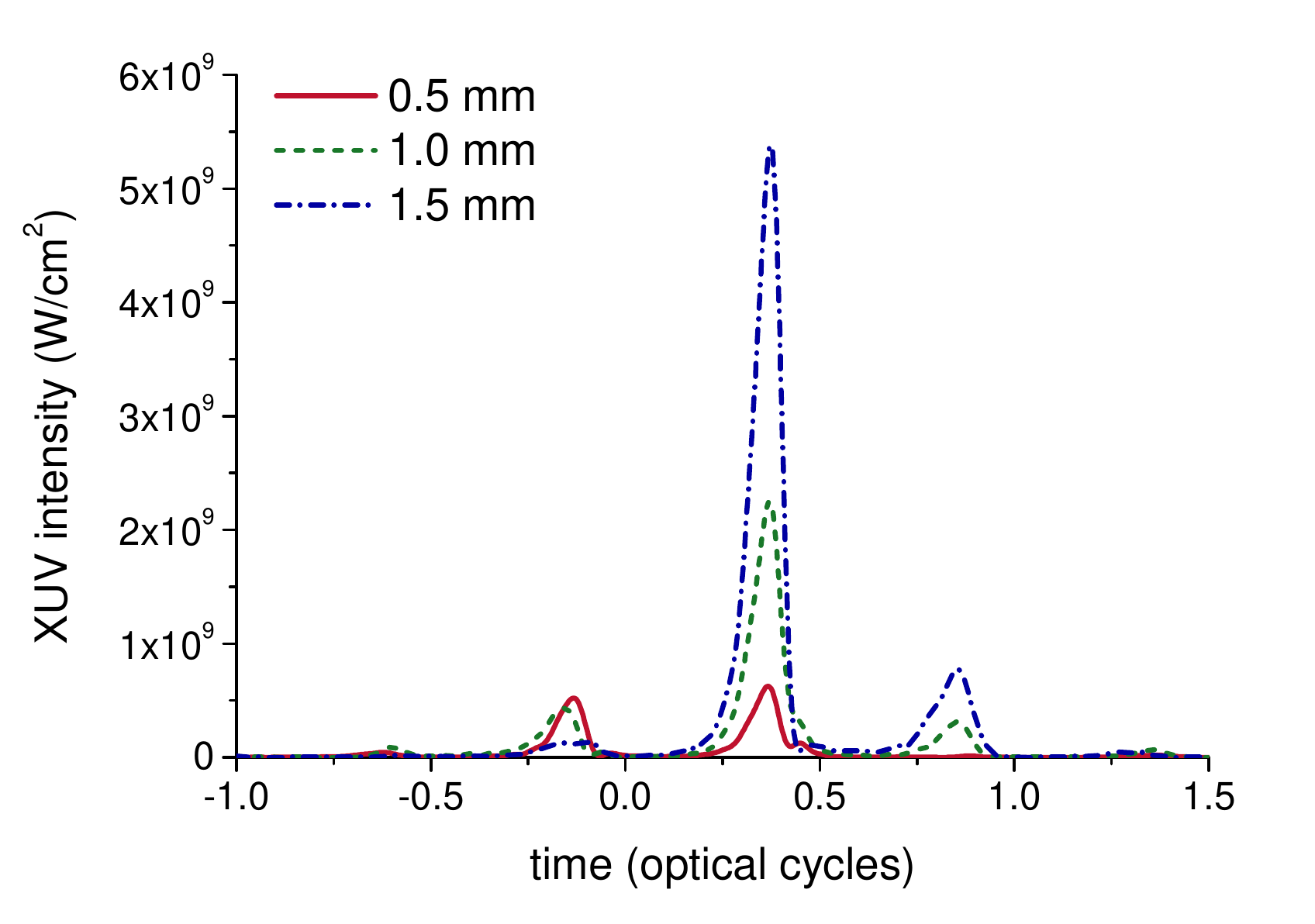}
\caption{Attosecond pulses calculated for different propagation distances. The laser pulse duration is \SI{10}{fs}, and its peak intensity is $\SI{2.6e14}{W/cm^2}$.} 
\label{Fig_attopulses}
\end{figure}

First, we look at the temporal behaviour of the emitted XUV. Fig.~\ref{Fig_attopulses} shows the intensity of XUV with frequency above H30 generated by a fundamental with \SI{10}{fs} duration (FWHM of intensity) as a function of time for three different propagation distances. One can see that after approximately \SI{1.5}{mm} of propagation an IAP is generated due to phase-matching gating. 

\begin{figure*}[]
\centering
\includegraphics[width=0.49\linewidth]{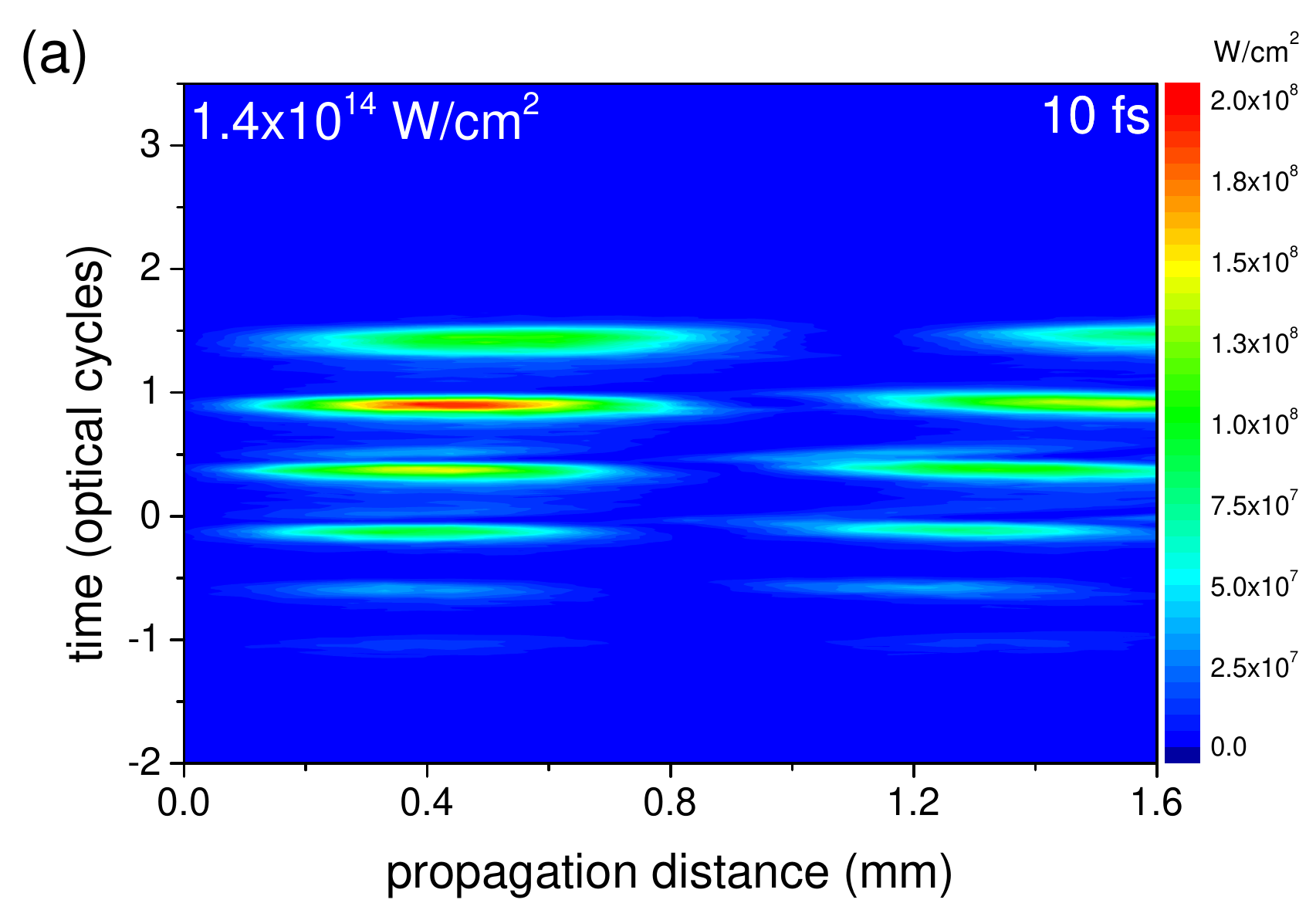}
\includegraphics[width=0.49\linewidth]{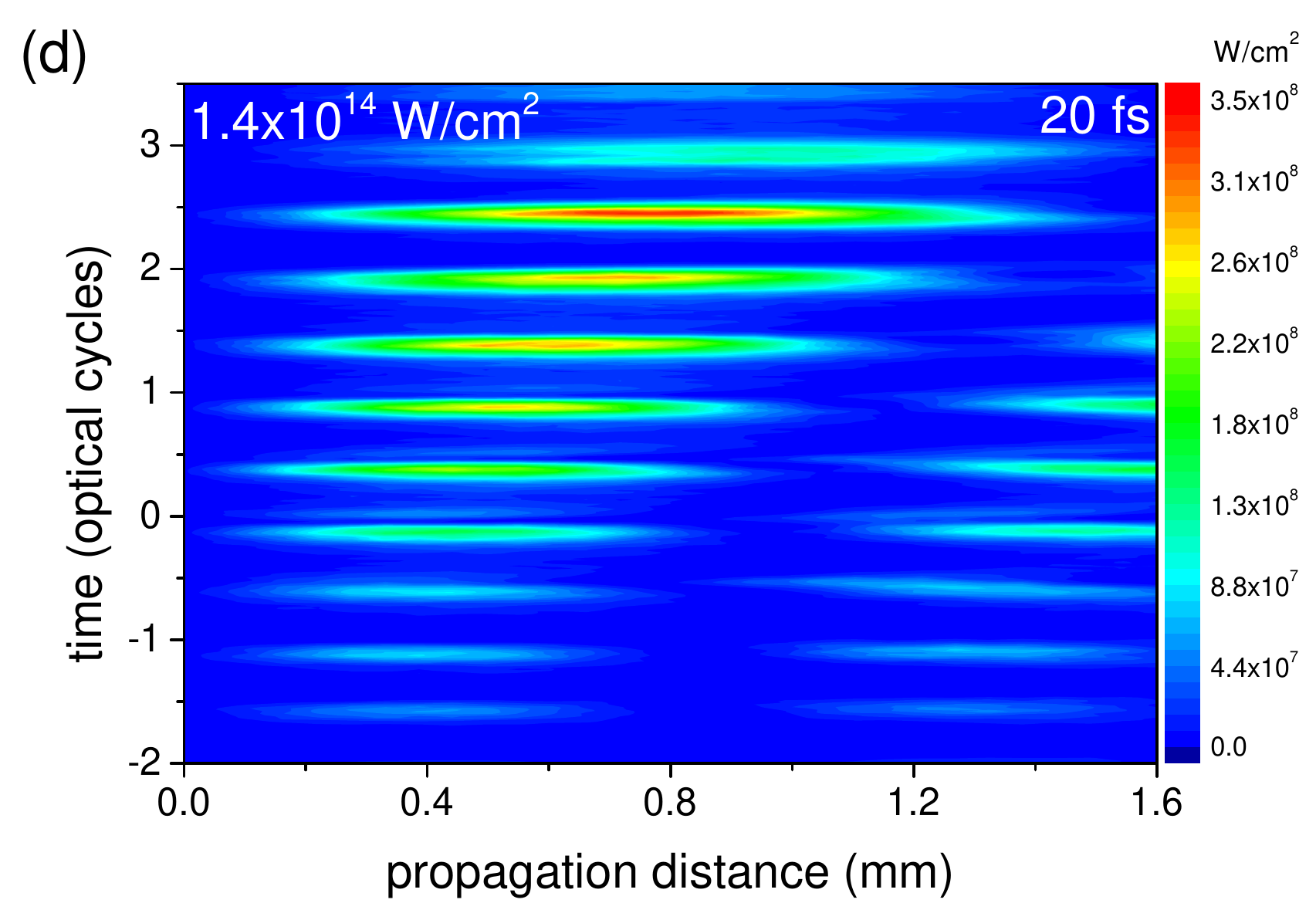}
\includegraphics[width=0.49\linewidth]{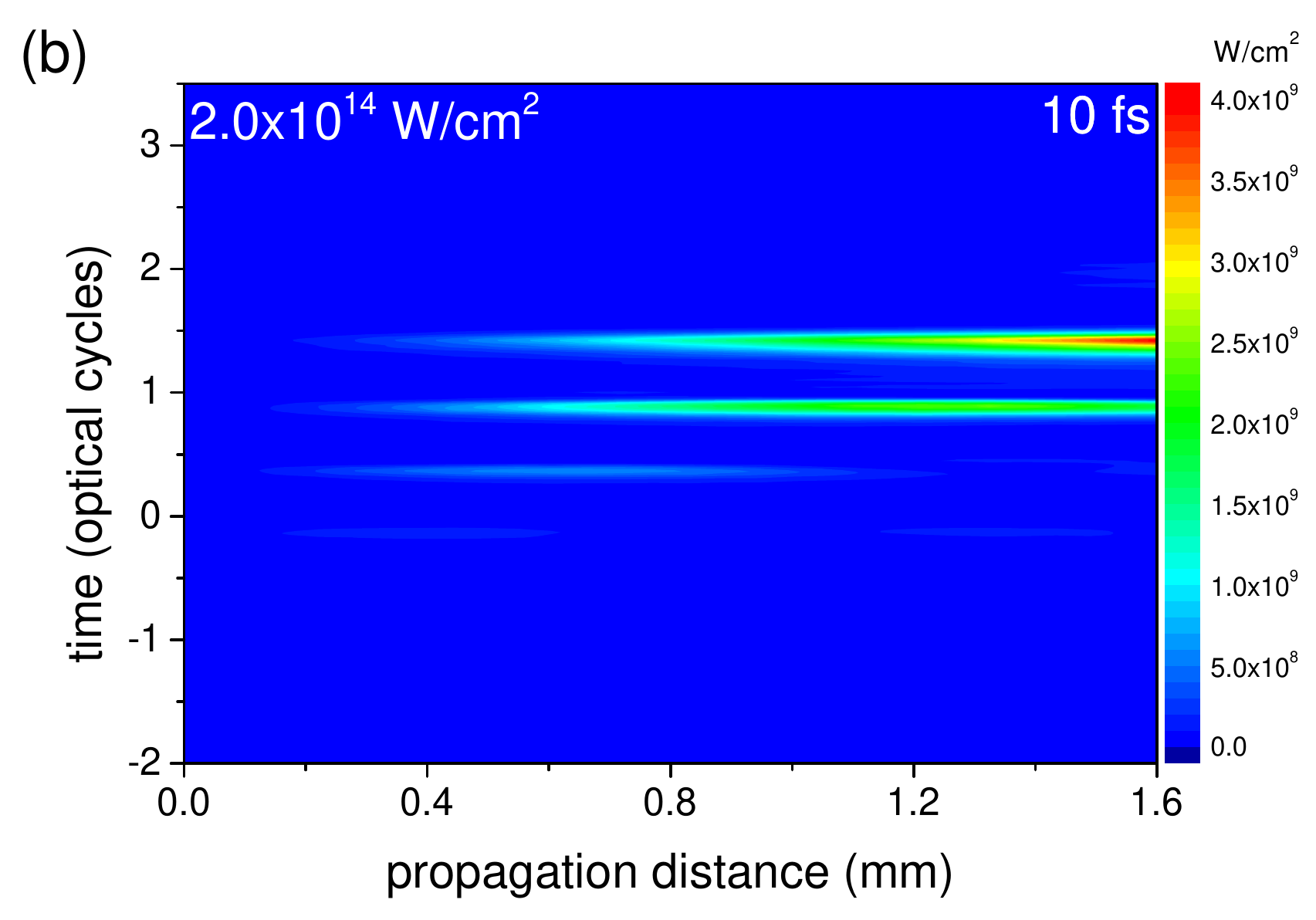}
\includegraphics[width=0.49\linewidth]{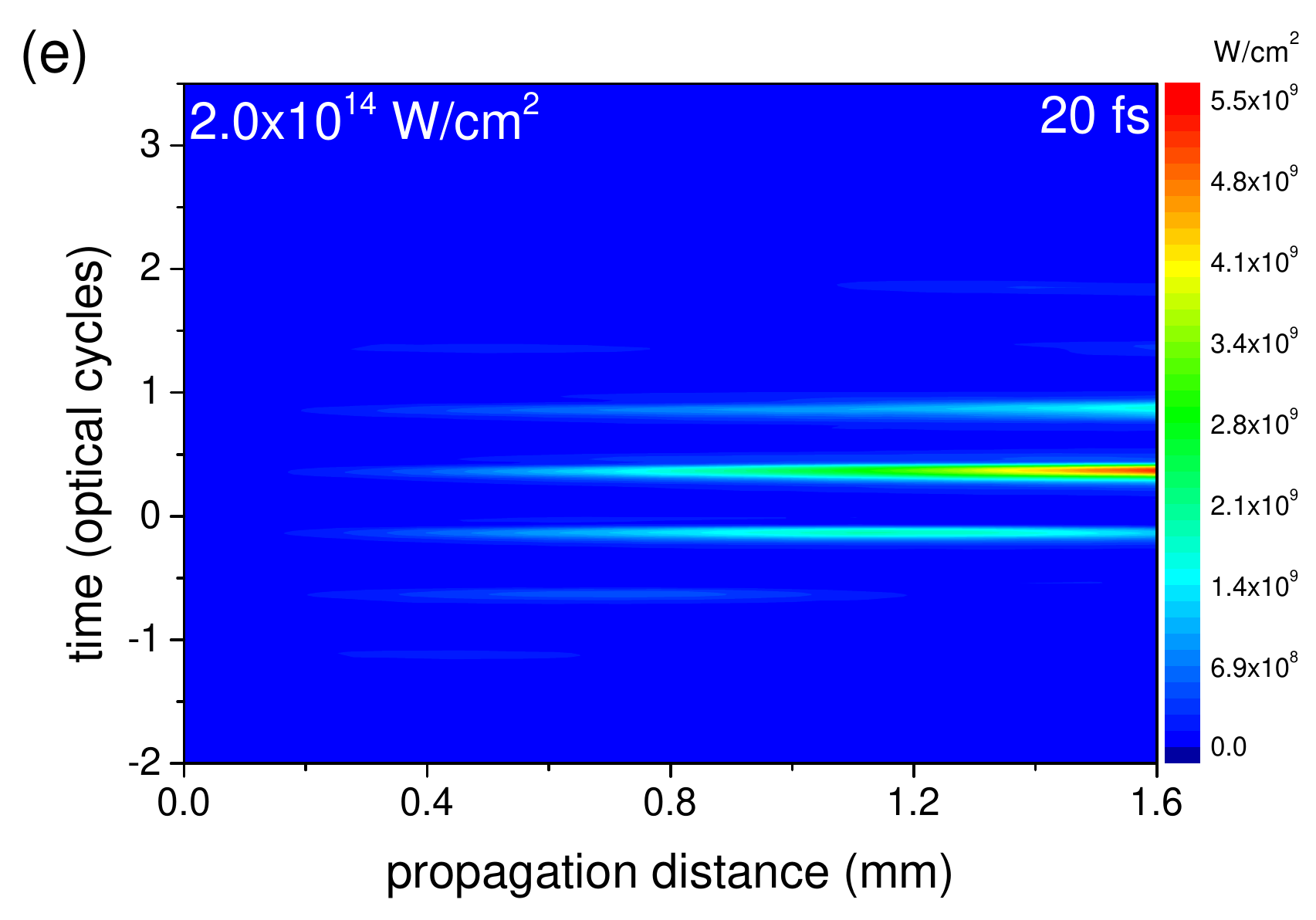}
\includegraphics[width=0.49\linewidth]{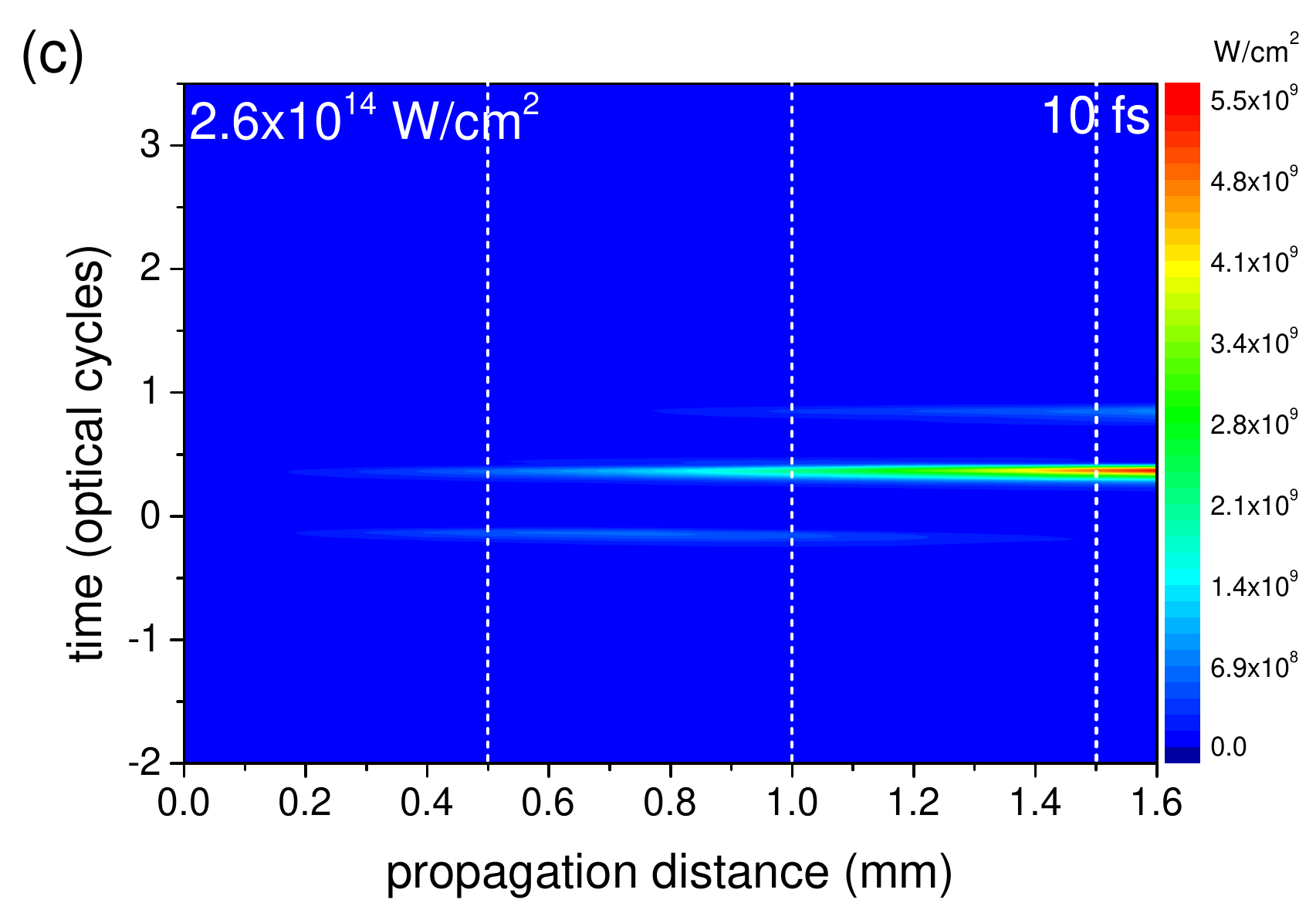}
\includegraphics[width=0.49\linewidth]{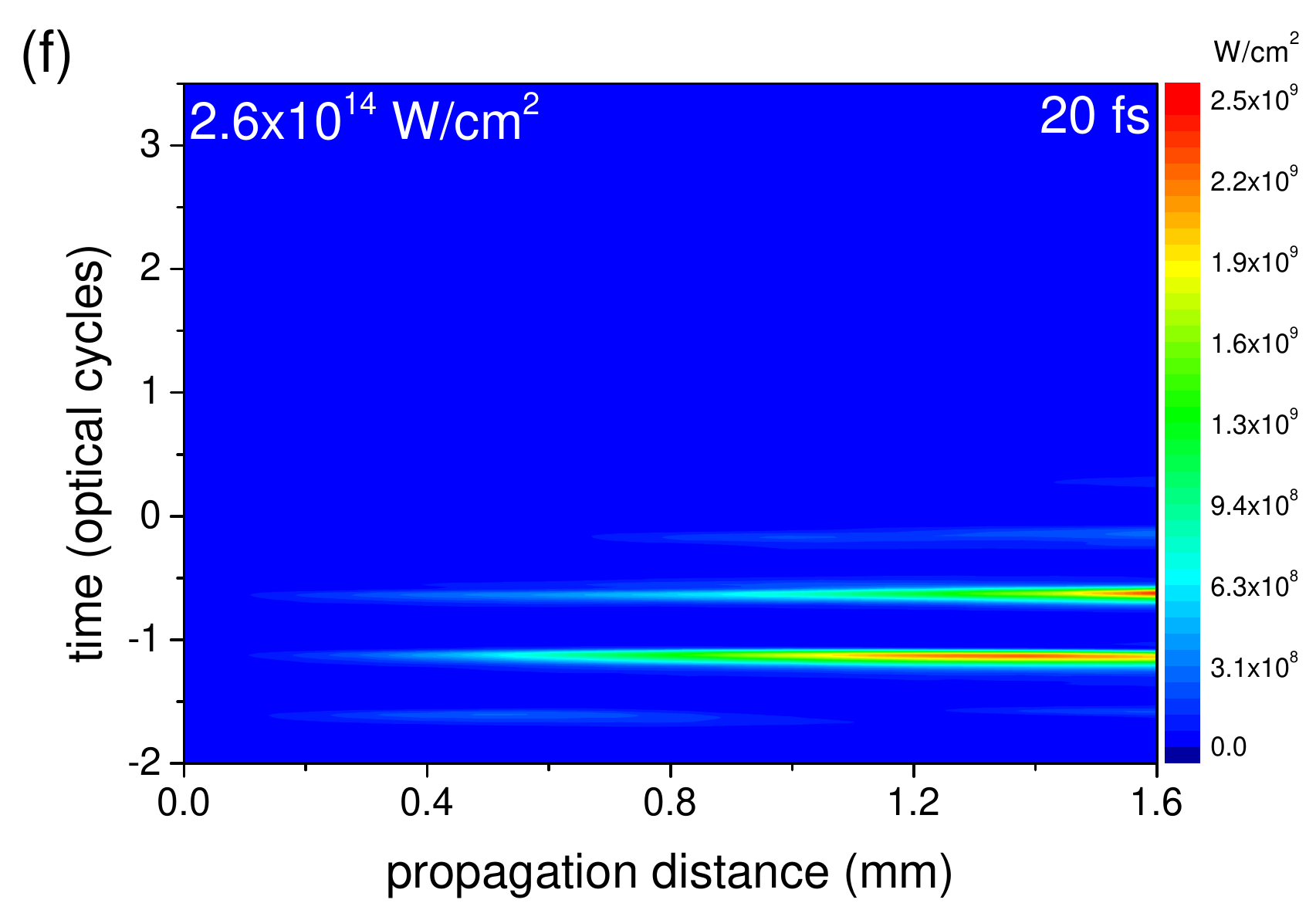}
\caption{XUV intensity as a function of time and propagation distance. The laser pulse duration is \SI{10}{fs}~(left column, or a, b, c) and \SI{20}{fs}~(right column, or d, e, f). The laser pulse peak intensity is $\SI{1.4e14}{W/cm^2}$~(upper row, or a, d), $\SI{2.0e14}{W/cm^2}$~(middle row, or b, e) and $\SI{2.6e14}{W/cm^2}$~(lower row, or c, f). Note different scales of the colour bars.}
\label{Fig_main}
\end{figure*}

To study the change of the temporal window of the phase-matching gating, we compare the temporal behaviour of the emitted XUV pulses, which are generated by laser pulses with different intensity and duration, during propagation. Fig.~\ref{Fig_main} shows the XUV intensity as a function of time and propagation distance for two different laser pulse durations and three different peak intensities. Each graph shows the XUV intensity with the photon energy above $I_p+2 U_p$, so this photon energy (and the harmonic order) is different for different laser intensities: H22 for the laser intensity $\SI{1.4e14}{W/cm^2}$, H26 for the laser intensity $\SI{2.0e14}{W/cm^2}$, and H30 for $\SI{2.6e14}{W/cm^2}$. Note that Fig.~\ref{Fig_attopulses} presents three slices shown in Fig.~\ref{Fig_main}(c) as vertical dashed lines.
One can see that with increasing laser intensity the number of emitted XUV pulses decreases, and the optimal conditions to produce IAPs in our case correspond to Fig.~\ref{Fig_main}(c).

\begin{figure}
\centering
\includegraphics[width=1.0\linewidth]{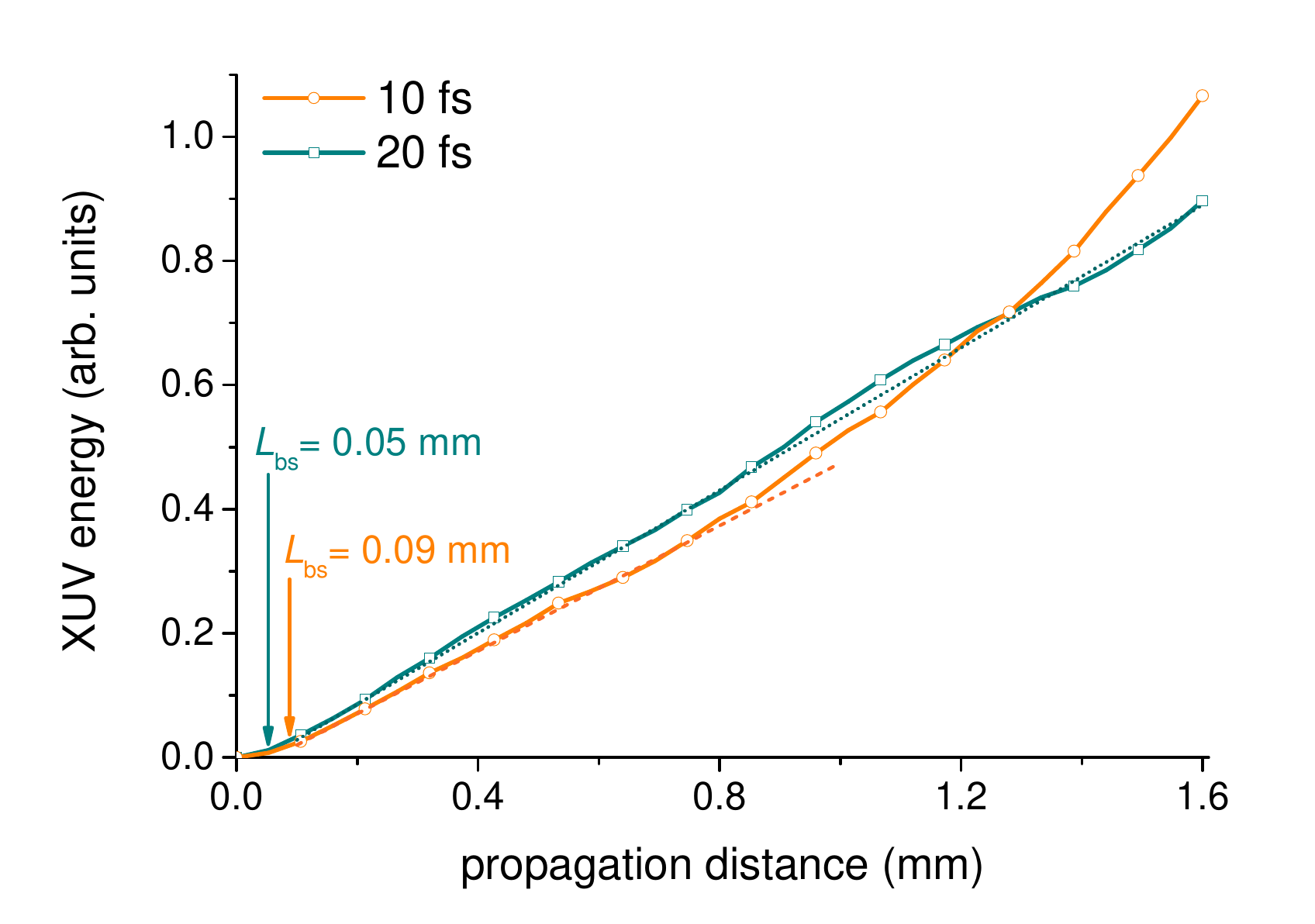}
\caption{XUV energy as a function of propagation distance for the laser pulse duration \SI{10}{fs}~(orange curve with circles) and \SI{20}{fs}~(cyan curve with squares). The dashed lines present the linear trend. The laser intensity is $\SI{2.6e14}{W/cm^2}$.} 
\label{Fig_En_L}
\end{figure}

Now let us consider the behaviour of the total XUV energy emitted as IAP or near-IAP for the most favourable intensity of the laser pulse and for \SI{10}{fs} and \SI{20}{fs} pulse durations. Fig.~\ref{Fig_En_L} presents the XUV energy~(the XUV intensity integrated over time) as a function of propagation distance. The dependence of the high-harmonic energy on the propagation distance was studied in~\cite{KhokhlovaStrelkov2020, Khokhlova2023}. One can see in Fig.~\ref{Fig_En_L} that initially the energy grows quadratically with the propagation distance. At the distance equal to the blue-shift length~$L_\mathrm{bs}$ (marked with arrows in the figure) the quadratic growth changes to a linear one.

The linear growth can be explained in the temporal domain as follows. XUV is generated mainly within the phase-matching window, where the dispersion of the neutral gas is compensated by the free-electron one, appearing due to photoionisation. The XUV intensity within the phase-matching window grows quadratically with propagation distance, but the temporal width of the window decreases as the inverse of the propagation distance (assuming linear temporal variation of the fundamental refractive index near the centre of the window $t_0$, we have the HHG phase mismatch $\Delta \varphi (t) \propto (t-t_0) L$; if the window's temporal width is $\tau$, then $|\Delta \varphi (t_0 \pm \tau)|= \pi$, thus $\tau \propto L^{-1}$), so the total XUV energy grows linearly.

\begin{figure}
\centering
\includegraphics[width=1.0\linewidth]{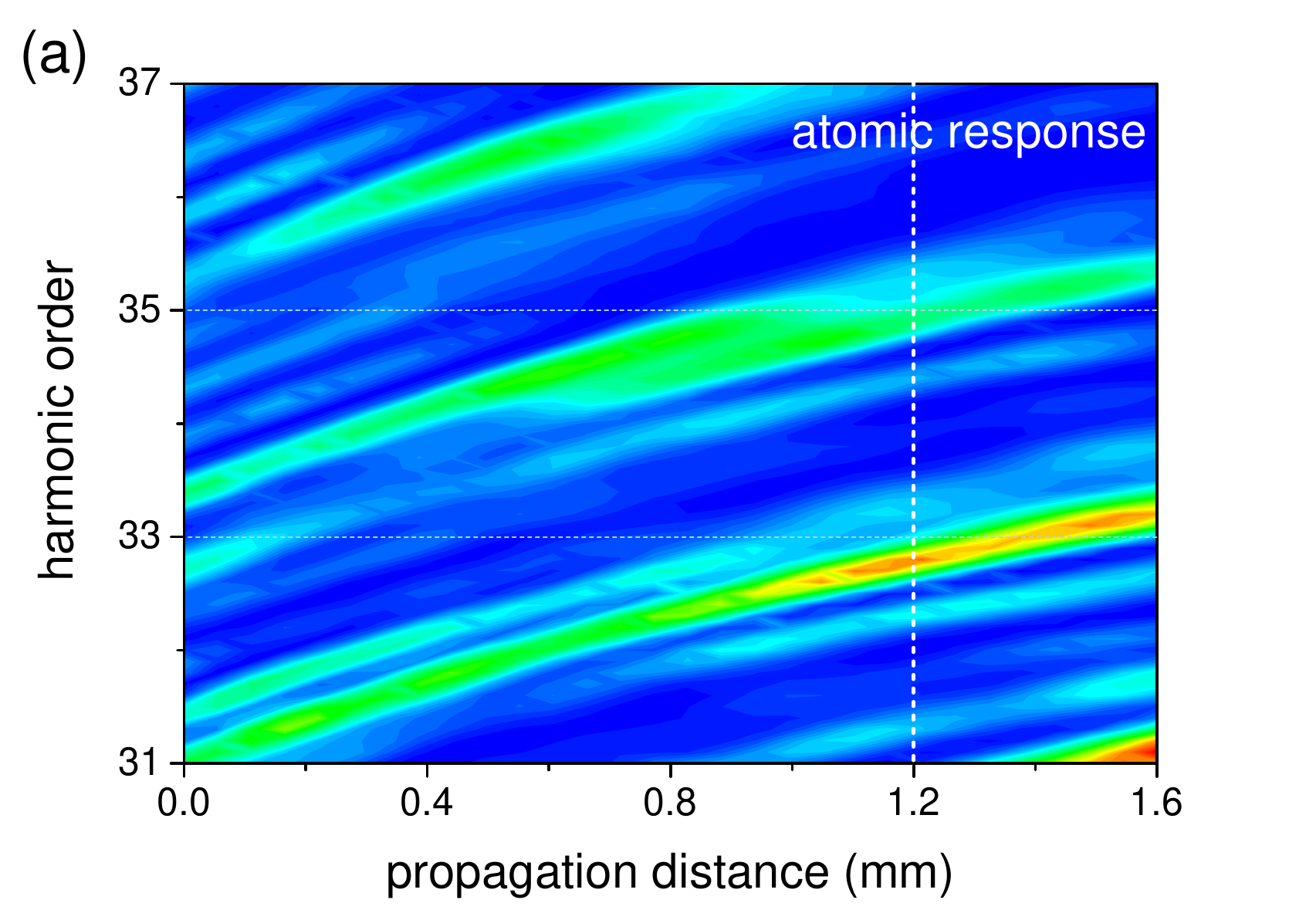}
\includegraphics[width=1.0\linewidth]{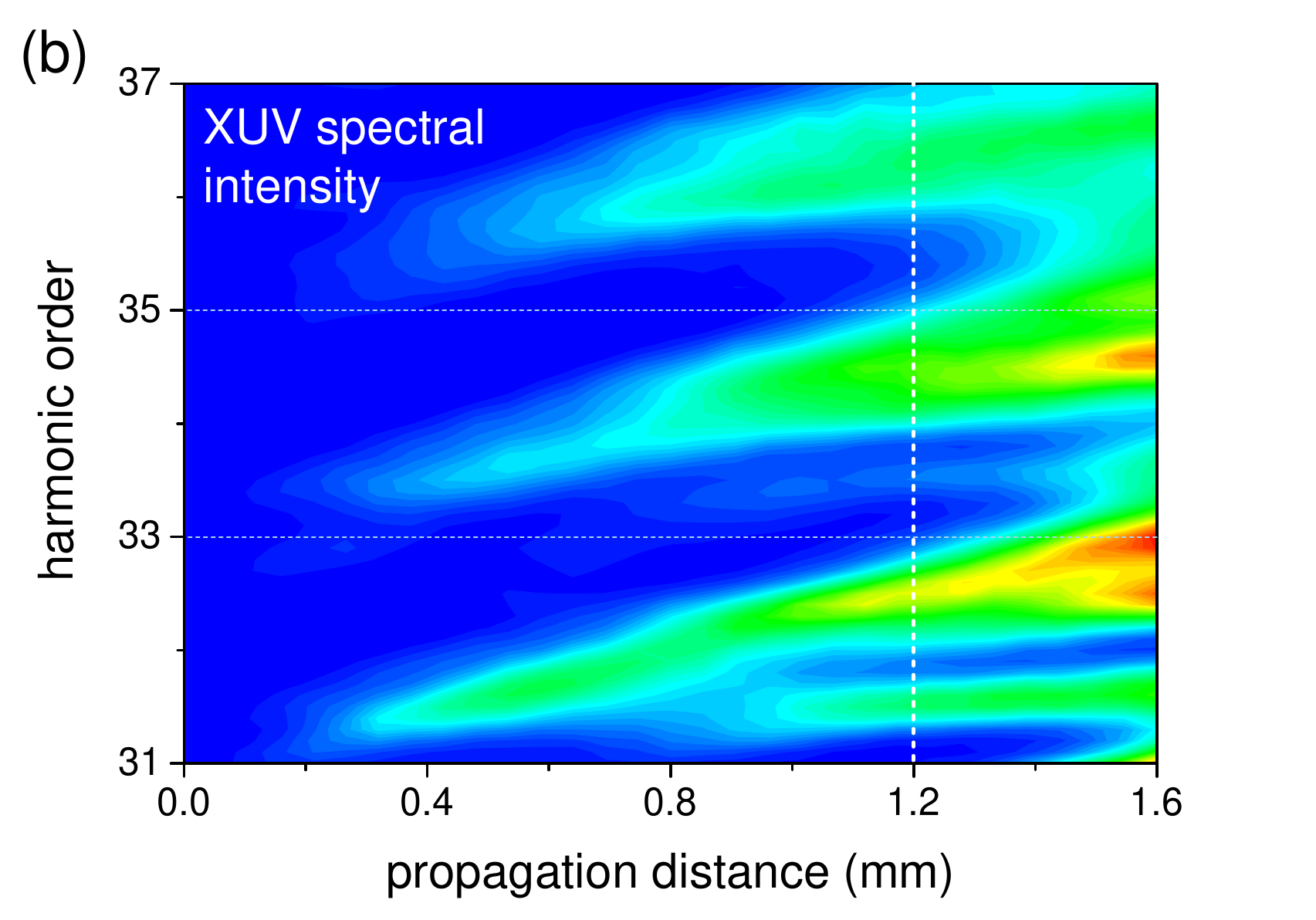}
\caption{Spectra of the atomic response~(a) and the propagated XUV field~(b) as functions of propagation distance. The laser pulse duration is \SI{10}{fs}, and its peak intensity is $\SI{2.6e14}{W/cm^2}$.} 
\label{Fig_sp_E_L}
\end{figure}

In the spectral domain this behaviour can be understood in the following way. The ionisation-induced blue shift of the fundamental leads to a blue shift of the harmonics in the spectrum of the microscopic polarisation induced by the total field, see Fig.~\ref{Fig_sp_E_L}(a). Thus, the spectral width of the propagated field of each harmonic, shown in Fig.~\ref{Fig_sp_E_L}(b), grows linearly with propagation distance, and its spectral intensity is approximately constant. As a result, the total XUV energy growth is linear.

However, in Fig.~\ref{Fig_En_L} one can see that for longer propagation distances, namely ones exceeding approximately \SI{1.2}{mm}, the XUV energy for the \SI{10}{fs} fundamental again grows faster than linearly, while for the \SI{20}{fs} fundamental this is not the case. This behaviour can be explained in the temporal domain as follows. For this relatively long propagation distance the phase-matching window becomes shorter than one half-cycle, so only one attosecond pulse, or IAP, is generated, see the dashed blue line in Fig.~\ref{Fig_attopulses}. Further decrease of the window's temporal duration almost does not influence this attosecond pulse (until the durations of the attosecond pulse and the window become comparable). The attosecond pulse intensity increases quadratically with propagation length, and so does the XUV energy. 

\begin{figure*}[t!]
\centering
\includegraphics[width=0.49\linewidth]{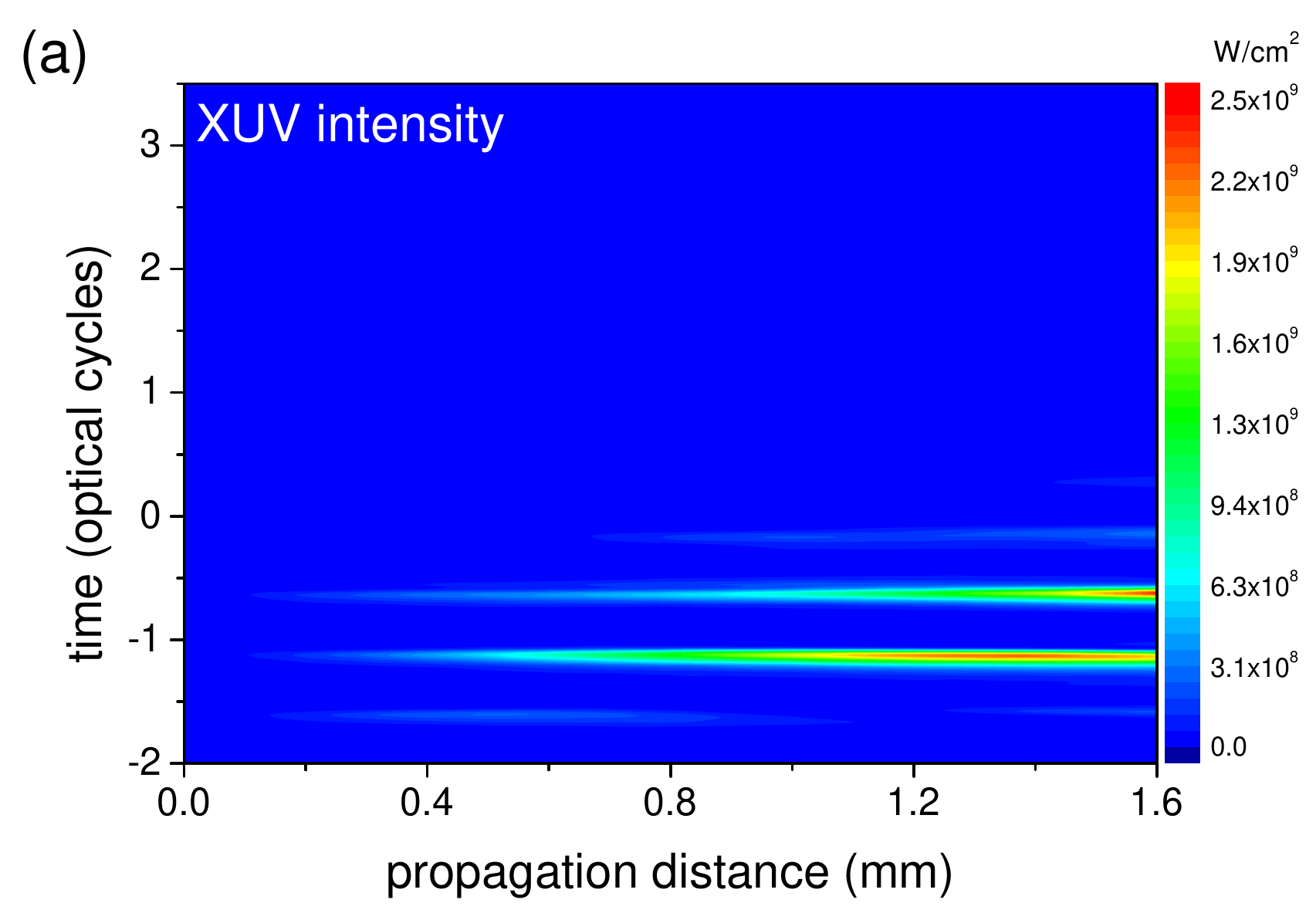}
\includegraphics[width=0.49\linewidth]{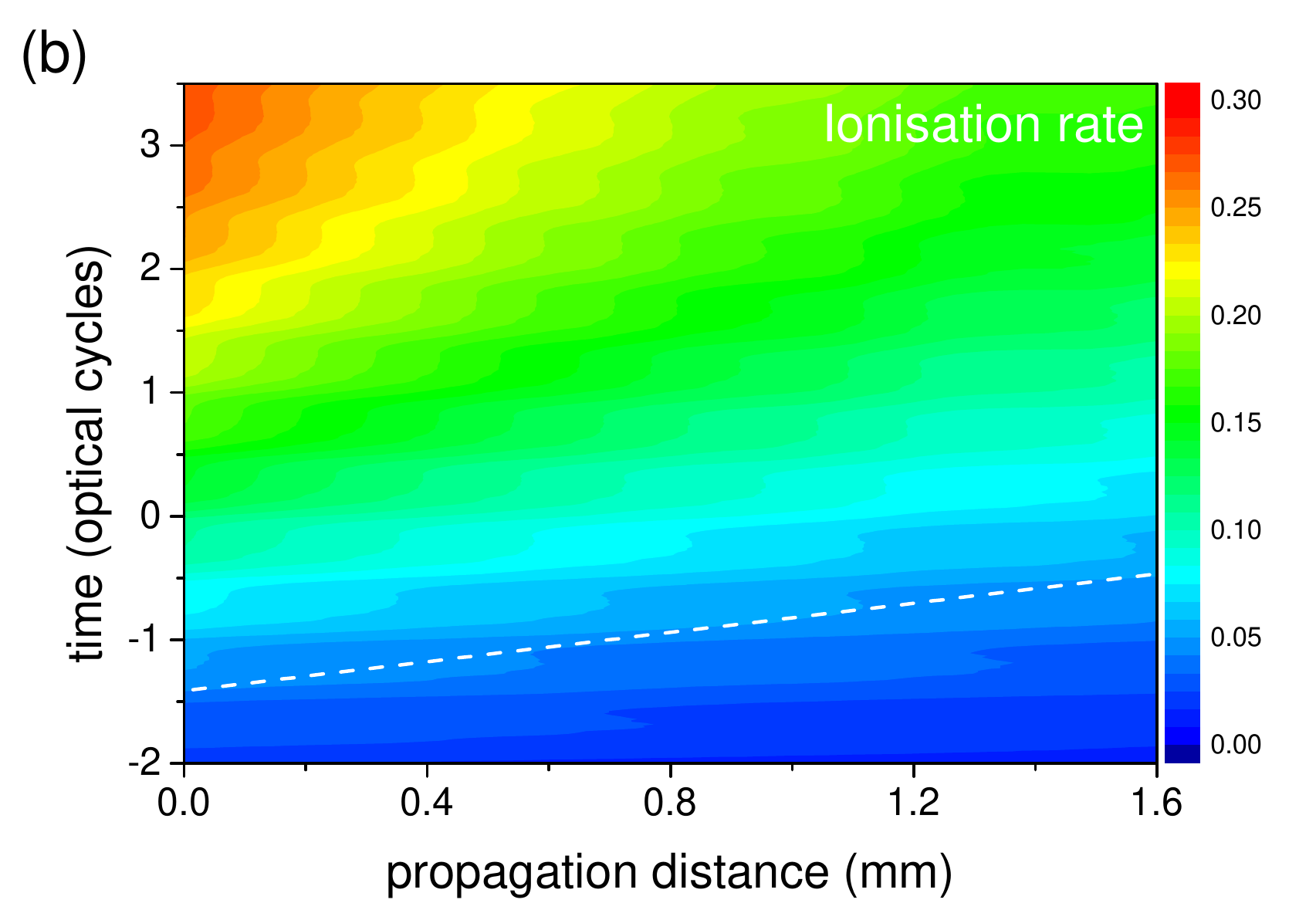}
\caption{XUV intensity~(a) and ionisation degree~(b) as functions of time and propagation distance. The dotted line shows the position of the phase-matching window for XUV generation. The laser pulse duration is 20 fs and its peak intensity is $2.6 \times 10^{14}$~W/cm$^2$.}
\label{Fig_ionization}
\end{figure*}

Now we look at the spectral domain vision of this behaviour. After approximately \SI{1.2}{mm} of propagation (marked in Fig.~\ref{Fig_sp_E_L} with vertical dashed lines) the frequency shift of the atomic response is close to twice the laser frequency, so the response at harmonic $q$ generated at short propagation distances superimposes with the response at harmonic $(q-2)$ generated at long ones, see Fig.~\ref{Fig_sp_E_L}. The constructive interference of these signals leads to the quadratic growth of the XUV signal with propagation. We would like to stress again that this change from linear growth to a quadratic one corresponds to the emission of a continuous spectrum, which results in the generation of an IAP. Note that in~\cite{KhokhlovaStrelkov2020, Khokhlova2023} we did not consider so long propagation distances leading to the overlap of the harmonic lines. 

For even longer propagation distances the XUV energy saturates and then decreases. In the temporal domain this is due to further narrowing of the phase-matching window, which becomes comparable with the attosecond pulse duration. In the spectral domain this situation corresponds to the accumulation of the phase mismatch of the contributions from different propagation lengths, and thus to their destructive interference. Note that our simulations underestimate the XUV absorption by the generating gas. In reality the saturation would originate mostly from the absorption of the XUV. 

Let us analyse the most favourable conditions for the IAP generation. From Fig.~\ref{Fig_main} we see that the IAP generation requires a short fundamental pulse and high intensity (this conclusion agrees with~\cite{phase-matching_gating}). The high intensity is required to achieve a rapid variation of the refractive index due to the gas photoionisation. However, at first glace, it seems that the same variation of the refractive index can be obtained for a longer pulse (\SI{20}{fs}) as well, so it is not transparent why IAP generation is not achieved using the \SI{20}{fs} fundamental. 

In Fig.~\ref{Fig_main} we can see that the phase-matching window shifts towards later times with propagation. To understand this feature in more detail, we present in Fig.~\ref{Fig_ionization}(a) the same results as in Fig.~\ref{Fig_main}(c), and in Fig.~\ref{Fig_ionization}(b) we present the ionisation dynamics corresponding to the same conditions. One can see that the optimal ionisation degree (the one which compensates the neutral-gas dispersion) is achieved later for longer propagation distances~--- this explains the temporal shift of the phase-matching window. This behaviour of the ionisation is explained as follows: the laser pulse temporarily spreads while propagating, so its peak intensity decreases. This leads to slower photoionisation at long propagation distances. Note that this feature was not taken into account in~\cite{phase-matching_gating}, where the laser pulse envelope was assumed to be the same for different propagation distances. Now we can conclude that the feasibility of IAP generation via phase-matching gating depends on the trade-off between the shortening of the phase-matching window and its temporal shift towards later times.

For shorter laser-pulse durations the laser intensity within the window is higher, so the photoionisation takes place faster near the window, and thus the shortening of the window occurs faster than for longer pulses. At first glance, it seems that the pulse spreading should also be more pronounced for the shorter pulse. This would be the case if the pulse spreading is given by the dispersion of the group velocity of a gas. However, this is not the case for our conditions. In our conditions the front of the pulse propagates in the neutral gas with its group velocity. The falling edge of the pulse propagates in the ionised gas with another group velocity. 

For the longer pulses the ionisation degree is higher, so the relative spreading is approximately the same for the short and the long pulses. Thus, for the shorter pulse the shortening of the window is faster and the temporal shift of the window is about the same as for the long pulse. As a result, the short laser pulse provides better conditions for the IAP generation. In Fig.~\ref{Fig_main} we see that the \SI{10}{fs} pulse provides such conditions, but the \SI{20}{fs} laser pulse does not. The absence of IPA generation for the \SI{20}{fs} laser pulse leads to the linear increase of the XUV intensity in Fig.~\ref{Fig_En_L}.

\section{Conclusions}
In this paper we study the production of an IAP due to transient phase-matching of HHG. For that, we numerically integrate the 1D propagation equation for the fundamental and generated fields with nonlinear polarisation calculated via 3D TDSE solution. The use of the 1D propagation equation is a reasonable approximation to describe HHG by a flat-top laser beam.  

We simulate the attosecond pulse generation by \SI{800}{nm} fundamental pulses with \SI{10}{fs} and \SI{20}{fs} durations and with different intensities in argon gas. We find that IAP generation via phase-matching gating is achievable using the intense \SI{10}{fs} laser pulse, and it is less feasible for the case of a \SI{20}{fs} laser pulse for the considered intensities. The reason of this discrepancy is the temporal spreading of the fundamental pulse during the propagation because of the medium photoionisation. This nonlinear effect is defined by the difference of the group velocities for the neutral and photoionised medium. The spreading leads to the temporal shift of the phase-matching window towards later times for longer propagation distances. 

Moreover, we study the behaviour of the XUV energy as a function of the propagation distance, and we find that for the \SI{20}{fs} pulse this dependence is linear, but for the \SI{10}{fs} one the onset of the IAP production corresponds to the change from linear to quadratic dependence of the XUV energy on the propagation distance. This feature can be used experimentally as a simple proxy method for the detection of IAP generation. Finally, it is important to note that the found dependencies on the propagation distance can be directly interpreted as dependencies on the medium pressure, as long as the 1D propagation picture is valid.

\section{Acknowledgements}
The study was funded by RSF (grant No 22-22-00242).

\bibliography{lit.bib}
\end{document}